\documentclass[aps,prl,floatfix,showpacs,tightenlines,twocolumn,superscriptaddress]{revtex4}

\usepackage{graphicx}
\usepackage{epsfig}

\begin{document}

\title{Weak nonlinearities: A new route to optical quantum computation}

\author{W. J. Munro}\email{bill.munro@hp.com}
\affiliation{National Institute of Informatics, 2-1-2 Hitotsubashi, Chiyoda-ku, Tokyo 101-8430, Japan}
\affiliation{Hewlett-Packard Laboratories, Filton Road, Stoke Gifford,
Bristol BS34 8QZ, United Kingdom}

\author{Kae Nemoto}\email{nemoto@nii.ac.jp}
\affiliation{National Institute of Informatics, 2-1-2 Hitotsubashi, Chiyoda-ku, Tokyo 101-8430, Japan}

\author{T. P. Spiller}\email{timothy.spiller@hp.com}
\affiliation{Hewlett-Packard Laboratories, Filton Road, Stoke Gifford,
Bristol BS34 8QZ, United Kingdom}

\date{\today}

\begin{abstract}
Quantum information processing (QIP) offers the promise of being able to do
things that we cannot do with conventional technology. Here we present a
new route for distributed optical QIP, based on generalized quantum
non-demolition measurements, providing a unified approach for quantum
communication and computing. Interactions between photons are generated
using weak non-linearities and intense laser fields---the use of such fields
provides for robust distribution of quantum information. Our approach requires only
a practical set of resources, and it uses these very efficiently.
Thus it promises to be extremely useful for
the first quantum technologies, based on scarce resources. Furthermore,
in the longer term this approach provides both options and scalability
for efficient many-qubit QIP.
\end{abstract}


\maketitle

\section{Introduction}
It is well known that using quantum information can
enable certain communication and computation tasks that cannot be performed with
conventional IT, or improve some that can\cite{Nielsen00}. Quantum computers (or smaller processors)
for instance can search faster and simulate better than their classical counterparts, and
factor large numbers efficiently\cite{Sho94,Gro96}. Quantum cryptography enables secure
communication, based on the laws of physics\cite{ref3}. There is now considerable research
activity devoted to finding
the best routes to realising new quantum information technologies. Here we
present a new paradigm for all-optical quantum information processing (QIP),
bringing together communication and computation in a single approach.
This has many appealing features and significant promise as a technology route:
\begin{itemize}
\item The approach is all-optical, so qubit interconversion is not a
prerequisite for combining communication and processing.
\item Quantum information is distributed robustly, using intense laser pulses.
\item Our approach only requires a practical set of physical resources. In particular,
we do not assume the existence of single photon sources and detectors, but
describe how these can be constructed from the underlying resources.
\item Moving on from quantum cryptography, the next technologies---based on
limited resources---are likely to involve few-qubit processing and some distribution
of quantum information. Our approach thus provides for this very efficiently.
\item Our approach provides adaptable building blocks, so the computational
approach is not fixed. For example gate- or measurement- or cluster-state-based
QIP can be performed, dependent upon the scale and application.
\item A longer term aim is for technologies based on many-qubit and distributed QIP.
Our approach gives an efficient and scalable route for this.
\end{itemize}

Optical QIP is already a very active research area. Knill, Laflamme and
Milburn (KLM) provided an important theoretical breakthrough, showing
that in principle universal quantum computation is possible
with linear optics, quantum gates effectively  being introduced through
photon bunching effects and
measurement\cite{KLM}. Furthermore, there have
been a number of recent impressive experimental demonstrations of the
building blocks for such optical QIP\cite{Pittman03,OBrien03,Gasparoni04}. However, despite all this,
gates in linear optical QIP are intrinsically probabilistic,
because they are based on photon bunching and measurement.
This means that the scheme is practically rather inefficient (in terms of photon resources)
to implement. Even with the application of cluster-state methods, on average over
100 photons are needed for a single two-qubit gate\cite{Nielsen04,Browne04}. In addition, all these photons
have to be identical enough for bunching effects to occur,
practically a very onerous requirement. So this road to actual
devices and technology looks tough.

An alternative method for realising a gate between two photons is to get them to
interact in a non-linear medium. An example is a cross-Kerr non-linearity,
with an interaction Hamiltonian $H_{ck}=\hbar \chi \hat n_a \hat n_p$
(where $\hat n_a=\hat a^{\dagger}_a \hat a_a$
and $\hat n_p=\hat a^{\dagger}_p \hat a_p$ are the photon number
operators for the two interacting modes and
$\chi$ is the interaction strength).
In order to produce sufficient
interaction for useful quantum gates directly between photons, strong
 non-linearities are needed, with $\theta \equiv \chi t \sim \pi$,
$t$ being the interaction time. Unfortunately, in practice such
non-linearities are not available. In effect, an approach is needed that can
``amplify'' the effect of the rather weaker non-linearities that are available
(with $\theta \ll 1$),to enable QIP\cite{munro03,grang98,nemoto04,barrett04,paris00}.

A key feature of our work is to do just this, by using intense coherent states
of light as a ``bus'' to mediate interactions between photon qubits. The
strength of the coherent states can offset the weakness of the non-linearities.
Two other important advantages arise from this approach. Firstly, photonic
qubits never talk to each other directly and so we do not rely on two-photon
effects between identical photons. Our photon qubits do not therefore have to be
perfectly indistinguishable, as they do in linear optical schemes.
Secondly, the coherent states used as bus
modes also provide for the natural communication and distribution of quantum information,
enabling gates between photonic qubits that never meet, and that are separated
by distances up to the scale over which
quantum communications work. As some quantum communication
tasks are more simply and robustly implemented with coherent states
(continuous variables (CV), as opposed to qubits), it is natural to combine both
qubit and CV resources in a truly distributed computing scheme.

Our distributed optical QIP scheme requires various fundamental physical
resources. Some of these are linear in nature, but for resource efficiency non-linear
elements are also necessary---for these we will specify
their nature and quantify the strength required.
In our proposal the fundamental physical resources required are:
\begin{enumerate}
\item Sources of large coherent states (which can clearly be filtered to make weak ones).
\item Highly efficient homodyne/heterodyne detection.
\item Standard linear optical elements, such as beam-splitters, phase shifters and
the ability to perform fast classical feed-forward).
\item Weak cross-Kerr non-linearities.
\end {enumerate}
With a view towards actual technology, this list is deliberately practical, so
it only includes coherent states of light---a standard optical resource---of the form
$|\alpha\rangle = e^{-|\alpha|^2/2}
\sum_n \alpha^n/\sqrt{n!} | n \rangle$, and weak non-linearities (with $\theta \ll 1$).
Quantum buffers or
storage for the qubits are not absolutely necessary; however, adding them to the list
renders actual implementations much simpler and even more efficient.

These core physical resources are all the elements needed to
perform any distributed single photon computation and communication task.
All the other necessary elements (photon sources, detectors and
gates) needed to enable the information processing can be created from
these core resources. Very importantly, no further hidden
source of non-linearities are needed in any of the elements, so it is
straightforward to cost the actual non-linear resources needed for any
given QIP task.

\section{Single Photon Sources and Detectors}

\begin{figure}[!htb]
\includegraphics[scale=0.4]{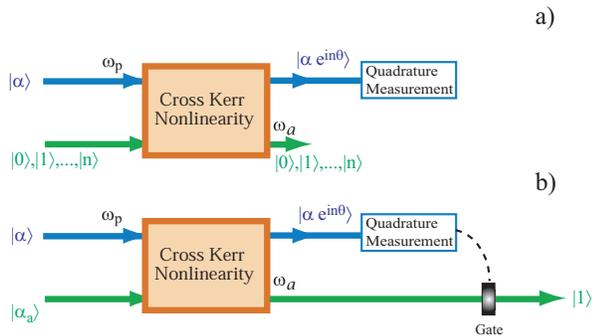}
\caption{Schematic diagram of a QND based single photon detector (a)
and a single photon source (b). For the QND based detector (a) the inputs
to the weak cross-Kerr material are Fock states $|n_a\rangle$ (with $n_a=0,1,$...)
in the signal mode labeled $a$ and a coherent state with real
amplitude $\alpha$ in the probe mode labeled $p$. The presence of photons
in mode $a$ causes a phase shift on the coherent state $|\alpha\rangle_p$
directly proportional to $n_a$ which can be determined with a
quadrature measurement. The single photon source (b) uses the QND detector from
(a) with a weak coherent state (mean photon number $|\alpha_a|^2 \sim 1$) input to the
signal mode. The homodyne measurement of the probe then allows the signal mode to be projected
into a definite number state. }
\label{fig-qnd-det}
\end{figure}

Our scheme embodies the actual quantum information in single photon qubits,
encoding into polarization of the photon ($|H\rangle, |V\rangle$),
or which of two paths it takes. (These encodings are equivalent,
and physical transformation between the
two is achieved through a linear polarizing beam-splitter.)
So single photons need to be created from the core physical resources detailed above.
We begin with
a brief discussion of how these core resources can be used to create
a high efficiency quantum non-demolition single photon detector\cite{Milburn84,Imoto85,munro03}, which can
be used to condition incoming photon states and thus
also serve as a single photon source.
In the quantum non-demolition detector (depicted schematically
in Fig. \ref{fig-qnd-det}a)  there are two optical modes, a signal
mode $a$ and a probe mode $p$. In our scheme the signal mode is some
superposition of Fock states and the probe
beam is an intense coherent state $|\alpha\rangle$. The signal and probe modes interact
via a weak cross-Kerr non-linearity that generates a unitary evolution
$U_{ck}=\exp \left[ i \theta \hat n_a \hat n_p\right]$, where $\theta$ is the total strength
of the non-linearity. If the signal mode is initially prepared in the state
$|\psi\rangle= c_0 |0 \rangle_a + c_1 |1 \rangle_a + c_2 |2\rangle_a$,
the cross-Kerr interaction causes the combined signal and probe system to evolve to
\begin{eqnarray}
c_0 |0 \rangle_a |\alpha\rangle_p
+ c_1  |1 \rangle_a |\alpha e^{i \theta }\rangle_p+ c_2  |2 \rangle_a |\alpha e^{2 i \theta }\rangle_p
\label{detoutputstate}
\end{eqnarray}
Now a highly efficient homodyne/heterodyne measurement of the probe field will
effectively project this onto some state of a chosen field quadrature
$x(\xi)=a_p e^{i \xi} + a^{\dagger}_p e^{-i \xi}$, and
the signal mode into a corresponding definite number state $|n\rangle_a$ \cite{Milburn82,munro03}. 
There is of course an error in the
discrimination of the $|n\rangle_a$ states due to the fact that the measured probe quadrature
probability distributions for the different $n$ have overlapping tails---the
phase shifted coherent states of the probe beam are not completely orthogonal to
$|\alpha\rangle_p$.
In terms of the measured quadrature value, the natural discrimination boundary for
adjacent $n$ values is the mid-point between the probability peaks. For the case of just
two peaks ($c_2 = 0$ in state (\ref{detoutputstate})), real $\alpha$ and measurement of
$x(\pi/2)$, the discrimination error is the
probability in the tails of the distributions on the wrong side of the mid-point, which is
$\frac{1}{2}{\rm erfc} \left[ |\alpha| \sin \theta/ \sqrt{2}\right]$.
With more peaks the total discrimination
error doesn't exceed $P_{\rm err}={\rm erfc} \left[ |\alpha| \sin \theta/ \sqrt{2}\right]$
which can be made small by ensuring that $\alpha \theta \gg 1$. In fact
for   $\alpha \theta \sim \pi$ this discrimination error $P_{\rm err}\sim 10^{-3}$,
which is likely to be smaller than other error and noise processes, so this approach
enables accurate non-absorbing photon measurement.

Next we focus on the generation of single
photons on demand. The QND detector is the required element for such a source. Consider
that the signal mode $a$ is now injected with a weak coherent state
$|\alpha_a \rangle$ ($\alpha_a$ real).
Such states are straightforward to prepare, and are in our list of core resources.
The QND detector with an appropriate
homodyne/heterodyne measurement of the probe mode will project the signal mode
into a photon number state $|n\rangle_a$. The value of $n$ is identified by the probe
measurement result. So production of a single photon state $|1\rangle_a$ is heralded
and occurs with probability
$P=\exp \left[-\alpha_a^2\right] \alpha_a^2$ which has a maximum value of
$1/e\sim 0.367$ when $\alpha_a\sim 1$. Several such driven sources can therefore give
a very high probability for the heralded generation of a single photon. This photon then
simply needs routing to where it is required.
Clearly the addition of a controlled quantum buffer or storage
device enables a probabilistic heralded single photon source to be turned into an ``on-demand''
single photon source. Essentially a probabilistic source can be used repeatedly until a
single photon is generated; it can then be held in a buffer until it is required.

\section{The distributed parity gate}

\begin{figure}[!htb]
\includegraphics[scale=0.4]{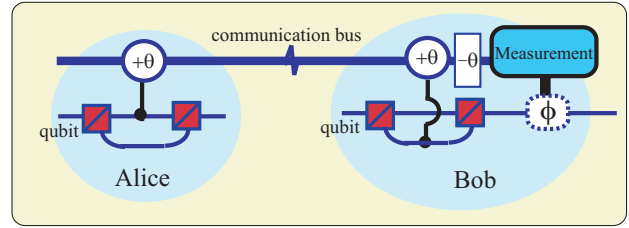}
\caption{Schematic diagram of a distributed two-qubit polarization QND
parity gate. The basic gate consists of two spatially separated photonic qubits,
an intense coherent state communication bus (the probe beam) and two
weak cross-Kerr non-linearities. The polarization encoded qubits are converted to
``which path'' qubits on polarizing beam-splitters and one path  of each qubit
interacts with a weak cross-Kerr non-linearity to induce a phase
shift $\theta$ on the communication bus. After both qubits have interacted with the
non-linear media a further linear phase shift of $-\theta$ is applied to the
communication bus, followed by a measurement. This projects the two-qubit state
into a subspace of even or odd parity. }
\label{fig-qnd-entangler}
\end{figure}

We have described how to produce both single photon sources and means to detect them
from our core physical resources.
Now we turn our attention to photon qubit interactions. There are obviously
a number of techniques or methods we could use to enable interaction between single photon
qubits, such as the approach of KLM\cite{KLM}.
However, we propose using weak cross-Kerr non-linearities and strong coherent states,
as this makes efficient use of resources from our core list and also enables a naturally
distributed quantum gate. Clearly a single weak non-linearity ($\theta \ll 1$) is not
sufficient to enable a maximally-entangling gate (such as CPhase) directly between two
single photon qubits---a great many non-linearities would be needed with this approach.
However, in conjunction with an intense probe beam just two weak non-linearities are sufficient
to implement a two-qubit parity gate\cite{nemoto04}.
Such a gate is illustrated in Fig. \ref{fig-qnd-entangler}.
The parity gate works as follows: consider that the two parties Alice and Bob
have the shared two polarization qubit state $\beta_0 |H  H\rangle_{ab} 
+ \beta_1 |H V\rangle_{ab}+ \beta_2 |V H\rangle_{ab}+ \beta_3 |V V\rangle_{ab}$
with each qubit at a different spatial location (and potentially stored or buffered). 
This state may be separable or entangled depending on whether Alice and Bob have interacted 
previously. The first action of the parity
gate is to entangle the probe beam $|\alpha \rangle_p$ with Alice's polarization qubit.
The $|H\rangle_a$ component of Alice's qubit causes a
$\theta$ phase on the probe beam while the  $|V\rangle_a$ component leaves the probe beam
unchanged. After this interaction the probe beam is transmitted to Bob,
who also uses a weak cross-Kerr nonlinearity $\theta$ to interact his
$|V\rangle_b$ component with the probe field. This entangles Bob's
qubit to the probe field and Alice's qubit. A linear phase
shift of $-\theta$ is then applied to the probe beam. The resulting three-party state
(Alice, Bob and probe) is
\begin{eqnarray}\label{ent-qnd}
|\Psi\rangle_{abp}&=&\left[ \beta_0 |H H \rangle_{ab} +\beta_3 |V V \rangle_{ab} \right] |\alpha\rangle_p   \\
&+& \beta_1 |H  V \rangle_{ab} |\alpha e^{i \theta}\rangle_p+\beta_2 |V H \rangle_{ab} |\alpha e^{-i \theta}\rangle_p.\nonumber
\end{eqnarray}
We notice immediately that dependent upon the type of measurement on the probe
beam $p$, the state of Alice's and Bob's system can be conditioned into a number
of distinct pieces. A parity measurement could be effected by distinguishing the probe state
$|\alpha\rangle_p$ from $|\alpha e^{i \theta}\rangle_p$ and
$|\alpha e^{-i \theta}\rangle_p$, but not (even in principle)
distinguishing $|\alpha e^{i \theta}\rangle_p$ from
$|\alpha e^{-i \theta}\rangle_p$.

There are various measurement strategies for the probe beam that can
realise this state conditioning. The simplest is a high efficiency
$x(0)$ quadrature homodyne measurement\cite{nemoto04,barrett04}. Whilst this is
in principle a straightforward  measurement to implement, it is by
no means optimal. A near-optimal measurement is preferable as this
enables the strength of the cross-Kerr nonlinearities used to be
near-minimal. To this end we propose performing a QND photon number
measurement on the probe beam. However, during the action of the
gate the probe beam has to have a large mean photon number in order
to ``amplify'' the effect of the weak non-linearities. Therefore,
before measurement the probe beam must be displaced by an amount
$D(-\alpha)$\footnote{A displacement operation $\hat D_p(-\alpha) =
\exp(\alpha^* \hat a_p - \alpha \hat a^{\dagger}_p)$ can be achieved
by inputting the probe beam to a highly reflective beam-splitter,
with a large (compared to the probe beam) coherent state on the
second input}. This results in the three mode state
\begin{eqnarray}
|\Psi\rangle_{abp}&=&\left[ \beta_0 |H H \rangle_{ab} +\beta_3 |V V \rangle_{ab} \right] |0\rangle_p \nonumber \\
&+& e^{-i \alpha^2 \sin \theta} \beta_1 |H  V \rangle_{ab} |\alpha (e^{i \theta}-1)\rangle_p  \nonumber \\
&+&e^{i \alpha^2 \sin \theta} \beta_2 |V H \rangle_{ab} |\alpha (e^{-i \theta}-1)\rangle_p \; .
\end{eqnarray}
Note that the $|H  V \rangle_{ab}$ and $|V H \rangle_{ab}$ amplitudes have picked up a phase shift
due to the displacement. These  phase shifts are unwanted but can be simply removed by 
static phase shifters (no feed-forward is required).. .
The overlap between the probe components of the even parity ($|0\rangle_p$) and odd
parity ($|\alpha (e^{\pm i \theta}-1)\rangle_p$) amplitudes is very small if
$\alpha \theta \sim \pi$. In this case the mean photon number of the
odd parity components is not large, being approximately  $\bar n_{p,odd} \sim 10$.
Hence a measurement of $n_p$ with a
QND photon number resolving detector cannot distinguish the
$|\alpha (e^{\pm i \theta}-1)\rangle_p$ components from each other, but can
distinguish these from the $|0\rangle_p$ components.
This results in Alice's and Bob's combined state being conditioned to
\begin{eqnarray}
|\Psi\rangle_{ab}=\left\{
\begin{array}{ll}
\beta_0 |H H \rangle_{ab} +\beta_3 |V V \rangle_{ab} & {\rm for}\;n_p = 0 \\
\beta_1 e^{i \phi(n_p)} | H V \rangle_{ab} +\beta_2 e^{- i \phi(n_p)}|V H \rangle_{ab} & {\rm for}\; n_p > 0
\end{array}\right.
\end{eqnarray}
where $\phi(n_p)= n_p \tan^{-1} \left[\cot(\theta/2)\right]$. For $\theta \ll 1$,  
$\phi(n_p)$ can be simplified to $\phi(n_p)=- n_p \frac{\pi}{2}$ which is in affect 
a sign flip for $n_p$ odd and no change for $n_p$ even. This phase shift $\phi(n_p)$ can then simply be 
eliminated via a classical feed-forward operation (a phase shift dependent on the result of the measurement) 
as  the QND measurement gives $n_p$ and $\theta$ is 
known. The classical feed-forward operation is needed  because a different operation is needed 
depending on whether $n_p$ is even or odd. In many computational circuits this feed-forward 
operation (determined by the result of the QND measurement) can be delayed and performed at the final measurement stage for the qubits.
We also need to point out that the error in discriminating the two components (even and odd parity states) 
is approximately $P_{err}\approx 10^{-4}$ for $\alpha \theta \sim \pi$. This is a near optimal measurement.

\subsection{Decoherence}

The distributed parity gate approach is clearly a very appealing method for the remote creation of
entangled states which, as we shall explain, can be extended to perform distributed
quantum computation. However, it is necessary to examine the
effects of noise and decoherence on this distributed approach to judge its real practicality.
One of the main sources of decoherence in the transmission of the probe
field from Alice to Bob is likely to be amplitude damping or photon loss in the channel. For instance 
if the channel is a fiber then photons from the probe beam will be absorbed as it 
is transmitted between the remote locations. 
There are a number of ways to treat this loss, with the simplest being to model
it via a beam-splitter of reflectively $\eta$ which discards
a portion of the probe beam while it is being transmitted between the
remote locations. It is assumed that $\eta$ does not vary with time, and that it can be
measured in advance through suitable test experiments, so its value is known.

Consider now the distributed parity gate, but with such loss on the probe beam.
Alice and Bob again prepare their qubits as
$|\Psi\rangle_a = c_+ |H \rangle_a+ c_- |V \rangle_a$ and
$|\Psi\rangle_b = d_+ |H \rangle_b+ d_- |V \rangle_b$ respectively. The parity gate is
performed as before, but now, crucially, with a slightly reduced displacement of
$D(-\alpha \sqrt{1-\eta^2})$ applied to the probe beam, due to the fact that some of this
beam has been lost during transmission and a phase correction $\eta^2 |\alpha|^2 \sin \theta$ 
applied to one of the two qubits. The beam loss
(the discarded output from the model beam-splitter) has to be traced over, which leaves
Alice and Bob with a mixed state. We focus on an example case, the regime with
$(1-\eta^2) |\alpha|^2 \left[1-\cos \theta\right]\sim 10$, which can certainly be attained
with physically reasonable parameters. Then the resulting mixed state is
\begin{eqnarray}\label{decohered}
\rho_{ab}(n_p = 0) &=& \lambda_+  |\Psi^+ \rangle_{ab} \langle \Psi^+ |_{ab} 
+ \lambda_- |\Psi^- \rangle_{ab} \langle \Psi^- |_{ab} \\
\rho_{ab}(n_p >0) &=& \lambda_+  |\Phi^+ \rangle_{ab} \langle \Phi^+ |_{ab} 
+ \lambda_- | \Phi^- \rangle_{ab} \langle \Phi^- |_{ab}
\end{eqnarray}
where $|\Psi^\pm \rangle_{ab} = c_\pm d_\pm |H H \rangle_{ab} \pm c_\mp d_\mp |V V \rangle_{ab}$,
$|\Phi^\pm \rangle_{ab} = c_\pm  d_\mp |H V \rangle_{ab} +c_\mp d_\pm |V H \rangle_{ab}$, and 
$\lambda_\pm = \frac{1}{2} \left(1 \pm e^{-\gamma}  \right)$ with
$\gamma= \eta^2 |\alpha|^2 \left[1-\cos \theta\right]$.  There are now two interesting points; 
first the effect of loss in the probe beam lead to mixing of the two qubits (rather than the loss of 
the qubit if it has been transmitted) and second the probe beam even with finite loss can project 
the qubits into parity states. We obviously want to operate in the regime of small  
$\gamma$ as we then effectively have the pure state $ |\Psi^+ \rangle_{ab}$, 
$ |\Phi^+ \rangle_{ab}$. However with moderate loss our two qubits can be heavily mixed but 
still retain some degree of entanglement which can be purified/distilled using standard techniques 
on the photonic qubits (the effect of loss on the probe causes a dephasing error in the 
photonic qubits). Next to be able to perform the parity measurement we require 
$(1-\eta^2) |\alpha|^2 \left[1-\cos \theta\right]\sim 10$ to have a low failure rate. 
If this cannot be achieved then the parity measurement gives one of three possible results: even parity, 
odd parity or indeterminate. In the indeterminate case the gate fails, but the information 
encoded in the photonic qubits is not necessarily lost and could be recovered. Alternatively 
the gate could be attempted again with fresh/re-prepared qubits.

There will obviously be other forms of error that one needs to deal with (for instance loss of photons in the 
qubits), but these can be deal with using the standard techniques available for linear optical 
quantum computation.

\section{Techniques of Computation}

We have indicated how to produce the components needed for optical
quantum information processing---sources of single photons,
detectors of single photons and two photon parity
measurements---from our list of fundamental resources. We now
consider the potential models for distributed optical processing.
There are a number of computational models available; here we
consider the standard gate-based schemes and computation by
measurement. This illustrates the flexibility of our approach, as
these two schemes can be realised from the same basic ingredients.

\subsection{Gate based computation}

\begin{figure}[!htb]
\includegraphics[scale=0.4]{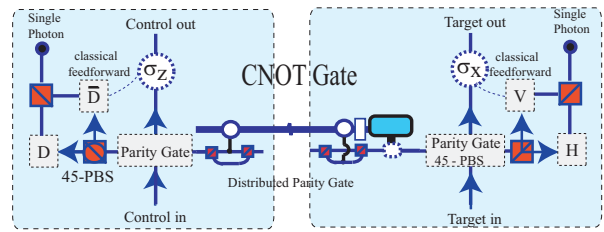}
\caption{Schematic diagram of a near deterministic CNOT comprising a
Bell state generator, two polarization qubit entangling gates (one
with PBS in the \{H,V\} basis and one with PBS in the \{H+V,H-V\}
basis), feed-forward elements and four single photon resolving QND
detectors.} \label{fig-qnd-franson}
\end{figure}

The standard gate-based model of universal computation requires
sources of single photons, single qubit rotations, an entangling
operation and projective measurements on these photons. For
polarization or which path encoded qubits, the single qubit
operations are essentially trivial and can be implemented using
linear elements; beam-splitters, phase shifter etc.. Historically
the tricky operation has been the entangling (two-qubit) gate. For
distributed processing we propose using coherent state bus modes,
which can suffer some loss, and thus avoid sending single photons
between remote locations. We propose an efficient distributed gate
based on an application of the linear optical CNOT ideas of
Franson\cite{Pittman01}. Our new gate is depicted in Figure
(\ref{fig-qnd-franson}) and comprises several local parity gates
plus a shared maximally entangled Bell state. This distributed
maximally entangled Bell state can be created by an application of
the parity gate to photonic qubits initially prepared as $ |H
\rangle+ |V \rangle$ and, crucially, just involves the transmission
of a coherent state between the two parties.
Consider the control and target qubits initially in the state
$\beta_0 |H  H\rangle_{ct} + \beta_1 |H V\rangle_{ct}+ \beta_2 |V
H\rangle_{ct}+ \beta_3 |V V\rangle_{ct}$. With a maximally entangled
ancilla Bell state $|HH\rangle_{ab} +|V V \rangle_{ab}$ the action
of the control location (left) parity gate and subsequent ancilla
qubit measurement projects and conditions the remaining three
photons into $\beta_0 |H H\rangle_{ct} |H\rangle_a + \beta_1 |H
V\rangle_{ct}|H\rangle_a+ \beta_2 |V H\rangle_{ct}|V\rangle_a+
\beta_3 |V V\rangle_{ct}|V\rangle_a$. Here a bit flip has been
applied on the remaining ancilla qubit for an odd parity gate
measurement result and a sign flip on the $|V\rangle_a$ ancilla
photon amplitude for a $\bar D$ ancilla qubit measurement result.
Application of the target location (right) parity gate, with the
standard PBS replaced with 45 PBS, to the target and remaining
ancilla qubit followed by a measurement on the ancilla qubit,
conditions the control and target qubits to
\begin{eqnarray}
\beta_0 |H  H\rangle_{ct}&+&\beta_1 |H V\rangle_{ct}+\beta_3 |V
H\rangle_{ct} +\beta_2 |V V\rangle_{ct}\;. \nonumber
\end{eqnarray}
Here an odd parity measurement result conditions a bit flip on the
ancilla qubit and a sign flip on the $|V \rangle$ amplitude of the
control qubit. Similarly if the QND measurement on the ancilla
yields $V$ then a bit flip is performed on the target qubit. The
action of all these measurements and corrections implements a CNOT
operation on the distributed initial state without transmitting
single photons between the remote locations. Using similar ideas it
is straightforward to show how three-qubit gates such as the Toffoli
and Fredkin gates can be constructed. It is obvious that these can
be constructed directly from their logical breakdown into two-qubit
gates. However, it is far more efficient to construct them directly
from the fundamental resources, using extensions of the primitives
employed in the parity and CNOT gates and only transmitting coherent
probe (bus) states between separated locations.

\subsection{Computation by projective measurements}

\begin{figure}[!htb]
\includegraphics[scale=0.65]{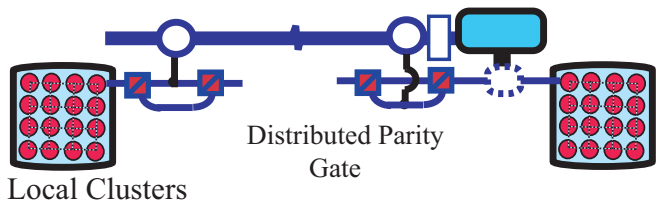}
\caption{Schematic diagram of a distributed cluster formed by joining two
local cluster states with the parity detector. }
\label{fig-qnd-cluster}
\end{figure}

An alternative approach to performing quantum computation is through
direct measurements applied to a collection of qubits. One of the
most well know examples is the cluster state approach\cite{Briegel01}. This requires
the creation of a specific type of entangled state---a cluster
state---of the qubits as an initial resource. The computation is
then accomplished by sets of single qubit measurements applied to
this entangled state. In linear optics this scheme can be
implemented using the Browne and Rudolph fusion techniques\cite{Browne04} 
and recently demonstrated in \cite{Walther05} . A
significant resource saving can be achieved by replacing the
beam-splitter based fusion gates with the parity gate.
This means that on average only one photon is needed to add one
photon to the cluster state (rather than 45 Bell pairs in the
original Browne and Rudolph scheme). The addition of qubits to the
cluster is thus near deterministic in nature. Also, because of the
distributed nature of the parity gate it is possible to envisage
situations in which local micro-clusters are generated and then
joined via distributed entangling operations (Fig
\ref{fig-qnd-cluster}). Clearly the single qubit measurements needed
to accomplish the computation once the cluster state is constructed
can be made within our proposed framework.

Other measurement-based approaches to quantum computation are also
available. Single qubit measurements first need the construction of
a cluster state. However, if any non-destructive two-qubit (or more)
projective measurements can be implemented these can facilitate
computation directly using the techniques of Nielsen and
Leung\cite{Nielsen03,Leung04}. A specific example is that of near-deterministic
Bell state measurements, which do not absorb the photon qubits. Such
measurements can be implemented in our framework by two applications
of the parity gate, with the second parity gate operating in a 45
degree rotated basis compared to the first. The implementation of
distributed projective measurements in an entangled basis is another
very appealing feature of our weak non-linearity approach.

Clearly the use of weak non-linearities enables several different
approaches to perform very efficient distributed optical
computation. The choice of approach will depend on many factors,
such as the scale and architecture of the processor, how distributed
it is, and the actual physical resources available. It is quite
possible that a hybrid approach, combining various computational
techniques, will be the most effective.

\section{Discussion}

We have presented a new paradigm for all-optical quantum processing,
based on generalized quantum non-demolition measurements. The
starting point is a practical list of resources (coherent states,
homodyne measurements, linear optical elements and weak
non-linearities) and we have shown how everything that is needed can
be built from these resources. As the approach is all-optical,
communication and computation blend together seamlessly with no need
for qubit inter-conversion. Distributed processing is enabled
naturally and, as coherent states mediate the quantum information
over distance, it is robust to photon loss. Furthermore, the use of
QND measurements makes our approach very efficient in terms of its
use of the fundamental resources.

The two basic QND building blocks in our approach are single photon
QND detectors and distributed two-qubit parity gates. The photon
detectors can be made very high fidelity and can also be used to
generate single photon qubit resources. The parity gate is
near-optimal in terms of its measurement scheme. Both detector and
parity gate require $\theta \alpha \sim \pi$; this
scaling is very important as it enables weak non-linearities
($\theta \sim 10^{-5}$) to be used with modest coherent states in
the probe or bus modes. The detectors and gates are not
deterministic, but near-deterministic. However, the errors arise due
to overlapping tails of different probe coherent states and these
can be made very small.

Our approach does not force a choice of computation scheme and
processor architecture; rather it provides building blocks which can
be put together to suit the task at hand. This, along with the very
high resource efficiency, makes our approach extremely flexible. In
the short term it provides a new way forward to distributed
few-qubit technologies based on scarce resources, and in the longer
term it provides a new approach to efficient, scalable optical
quantum computing either between distinct nodes or within a node. 

\noindent {\em Acknowledgments}: We thank Rod van Meter, S. Barrett, R. Beausoleil, S. L. Braunstein, P. Kok
and G. J. Milburn  for numerous valuable discussions. This work was supported in 
part by Japanese JSPS, MPHPT, and Asahi-Glass research grants and the European Project
RAMBOQ.

\end{document}